Road Safety and Simulation 2026 – RSS2026

# Virtual Reality-Simulated Interaction Between Micro-Mobility Vehicles and Pedestrians: A Biomechanical Analysis of Human Gait and Movement Responses

Sahan Siriwardena[a], Omar Aboelrous[b], Anas Madkoor[b], Mohamed Elansari[b], Aiman Alhetari[b], Jassim Al Jufairi[d], Mohammed Al-Kuwari[d], Qinaat Hussain[b], Charitha Dias[c], Ajith Sominanda Herath[d]*

[a]*Department of Electrical and Electronic Engineering, Faculty of Engineering, University of Peradeniya, 20400, Sri Lanka.*
[b]*College of Engineering, Qatar University, P.O. box 2713, Qatar.*
[c]*Department of Civil and Environmental Engineering, Khalifa University. UAE.*
[d]*Department of Basic Medical Sciences, College of Medicine, QU Health. Qatar University, P.O. box 2713, Qatar.*

**Abstract**

Pedestrian walking is a fundamental activity of daily living and a key component of first and last-mile urban mobility. The rapid adoption of e-scooters has increased pedestrian-vehicle interactions on shared sidewalks and crossings, raising collision risks. However, most previous studies have relied on trajectory-based observations, providing limited insight into biomechanical gait responses. This study investigated pedestrian gait adaptations during simulated e-scooter interactions using immersive virtual reality (VR) and markerless pose estimation. Twelve healthy male university students (21-23 years) completed four VR walking scenarios: normal walking, e-scooter encounters at 10-25 km/h, crossing encounters, and near-crash encounters. Sagittal-plane videos were analyzed using the OpenPose 25-point model. Step length, gait cycle time, walking velocity, stance and swing phases, and lower-limb joint trajectories were extracted using Kinovea and custom JSON-based analysis tools. Statistical analyses included ANOVA, MANOVA, and non-parametric tests Crossing and near-crash scenarios significantly reduced step length ($p<0.001$), from 226.5 cm during normal walking to 204.7 cm during near-crash simulations. Although gait velocity and timing were not significantly affected, participants consistently exhibited shorter stance phases, longer swing phases, and restricted knee motion during stressful encounters, indicating reflexive gait adaptations to perceived collision risk. These findings demonstrate that immersive VR combined with markerless pose estimation effectively quantifies pedestrian biomechanical responses to micro-mobility interactions. Gait adaptations identified in this study may serve as sensitive indicators of collision risk and support the development of proactive pedestrian safety measures and intelligent micro-mobility control systems.

* Corresponding author.
*E-mail address:* ajithsomi@qu.edu.qa

## 1. Introduction

Walking is a fundamental activity of daily living and a critical first and last mile travel mode in urban environments, where pedestrians routinely interact with vehicles on crosswalks and sidewalks. Existing research on pedestrian walking has primarily relied on field observations and trajectory-level analyses, with limited attention to detailed biomechanical gait responses, particularly during encounters with micro-mobility vehicles such as bicycles and e-scooters (Ni et al. 2016). Worldwide, the increasing use of e-scooters, the expansion of shared sidewalks and bicycle lanes, and reported e-scooter-related injuries highlight the need for context-specific data on pedestrian–micro-mobility interactions (Fu et al. 2019).

Interception of moving targets is a complex behavior that relies on tightly integrated perceptual and motor processes in the brain, engaging a distributed network of brain areas and recruitment of skeletal muscles if necessary (Merchant and Georgopoulos 2006). Thus, e-scooter encounters trigger a complex vasomotor reflex that will significantly alter the pattern of pedestrian movements. One of the objective ways to capture this response is to observe and measure gait biomechanics. This study uses virtual reality (VR) to characterize pedestrians' biomechanical and stress responses during such interactions in a controlled, realistic traffic environment.

## 2. Methodology

### *2.1. Participants*

A cohort of randomly selected healthy university students with written informed consent was initially screened for eligibility. Inclusion criteria comprised normal walking, absence of any previous or current musculoskeletal or neuromuscular disorders, and adequate tolerance and compliance with immersive virtual reality (VR) visual conditions and equipment. Of these, 20 participated in the study. All participants were given detailed instructions, followed by a demonstration of VR immersion and walking, along with an assurance of safety. A post-experimental questionnaire inquired about VR-related discomfort, dizziness, or a feeling of sickness. Of the total participants, the dataset of 12 students was found to be complete and suitable for final analysis. The experiment was conducted at the Qatar Transportation and Traffic Safety Center VR Laboratory following approval from the Institutional Review Board (QU-IRB 280/2024-EA). Data collection was carried out between January 2025 and December 2025.

### *2.2. VR Simulation and 2D Video Recording*

A 360° action camera (Insta360 X4, Insta360, Shenzhen, China) was used to record real VR immersive scenarios, as shown in Fig. 1.

- Non-stressful pedestrian-only walking
- E-scooter encounters at 10 to 25 km/h
- E-scooter crossing encounters
- E-scooter crash-in encounter

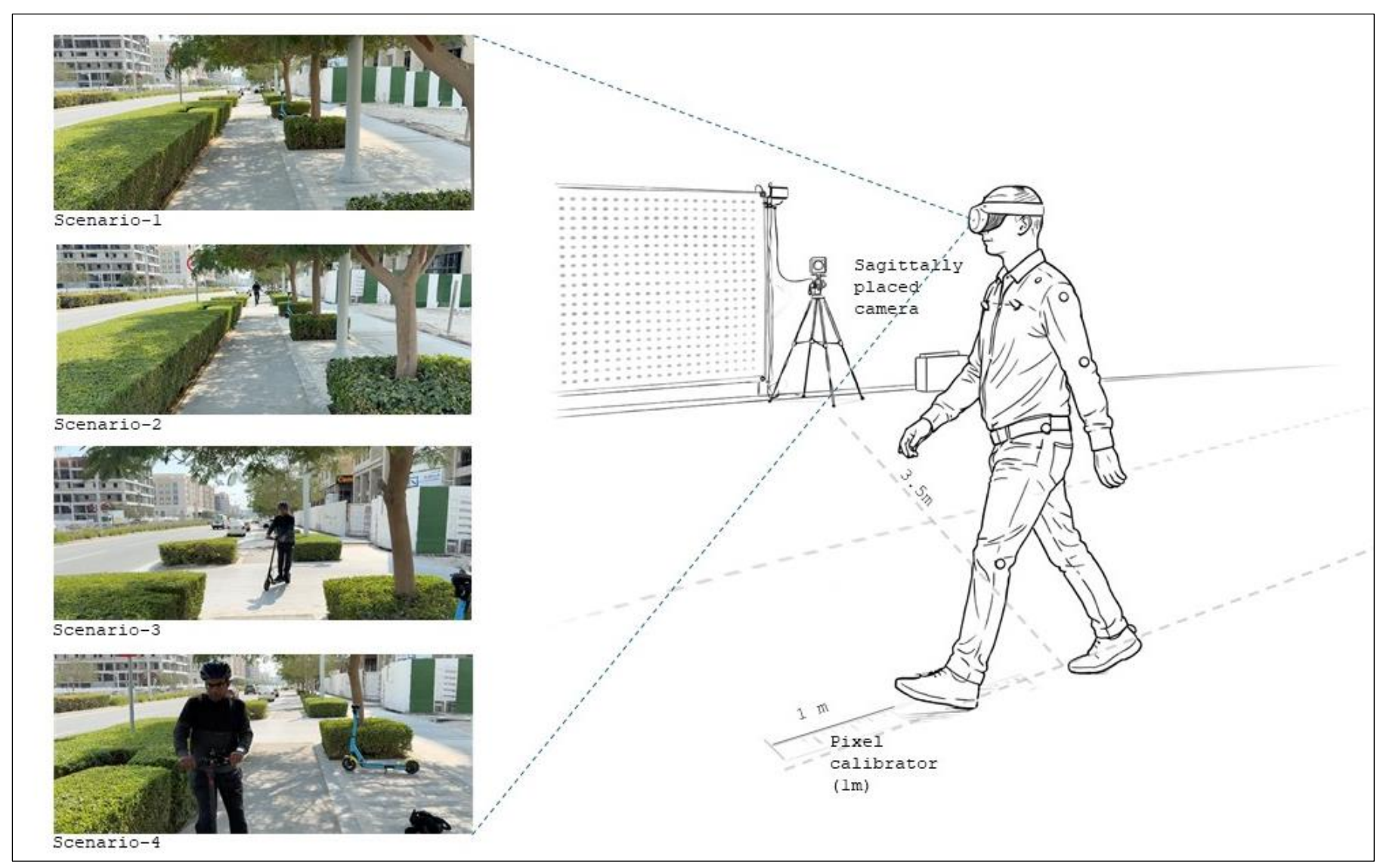

Fig. 1. Experimental VR scenarios and 2D video-based motion capture setup. Four VR scenarios include (1) baseline walking, (2) e-scooter encounters at 10–25 km/h, (3) crossing interactions, and (4) near-collision events.

The 360° action camera was also moved forward at a normal walking speed to mimic a pedestrian walking along the same walkway on the right side. The processed video stimuli were presented via a head-mounted virtual reality system (Meta Quest 3, Meta Platforms, Menlo Park, CA, USA) to induce immersive and controlled visual environments during gait trials.

Sagittal-plane video was captured using an action camera (GoPro HERO 12 Black, GoPro, San Mateo, CA, USA) at a resolution of 1080p (1920 × 1080) and a frame rate of 30 fps, mounted on a tripod at hip height (100 cm) and positioned 3.5 m from the midpoint of the walking stretch (5–6 m) in the VR laboratory (Fig. 1). The participant started walking from right to left according to the text and voice instructions embedded into the VR visuals. Each scenario was displayed for 5 seconds, and the tests were repeated whenever necessary.

### *2.3. Gait Analysis*

#### *2.3.1 Spatiotemporal gait parameters*

Sagittal-plane gait 2D videos were analyzed using Kinovea software (version 6.5) to extract spatiotemporal gait parameters. Two gait cycles (GC) were analyzed; defined as starting with the first left heel strike observed in the sagittal-plane video and continuing for two consecutive gait cycles, ending with the subsequent left heel strike (Fig. 2). Following spatiotemporal parameters were manually estimated, i.e., distance per 2 gait cycles, time, gait velocity, step time, stance, and swing phase time (Pulido-Valdeolivas et al. 2013, Sethi et al. 2022).

#### *2.3.2 Joint kinematics*

##### *2.3.2.1 2D Video Processing and Pose Estimation*

Each scenario-based sagittal-plane videos were used for marker less pose estimation using OpenPose (Cao et al. 2021) multi-person 2D pose detection pipeline that detects Body key points (e.g., BODY_25; enabled estimation of lower-limb, upper-limb, trunk, and head joint positions), Confidence maps and Part Affinity Fields PAFs). Of these, we mainly focused on left lower limb gait kinematics.

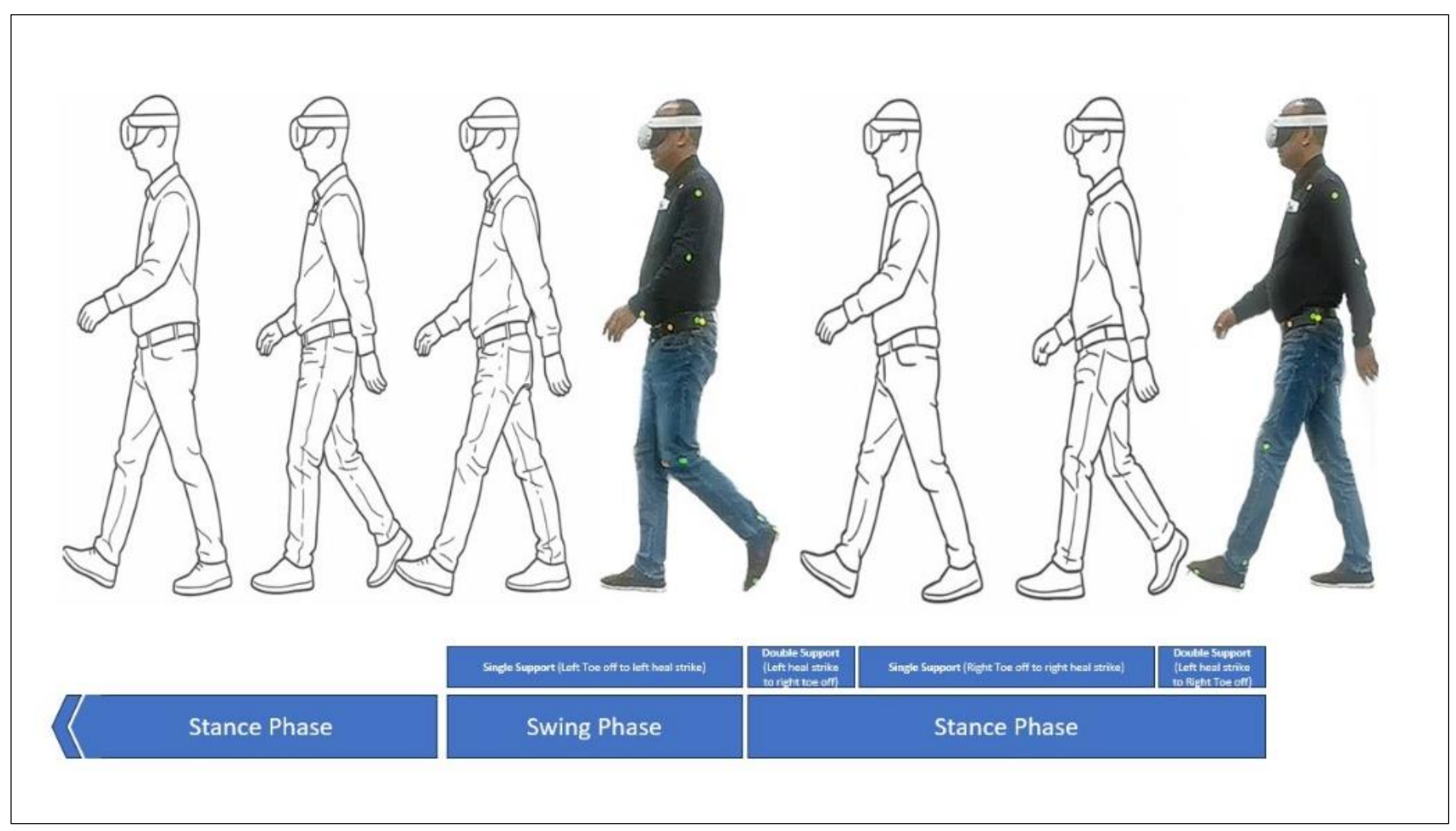

Fig. 2. Gait phases

For each processed video, OpenPose generated output files in JavaScript Object Notation (JSON) format, containing the x–y coordinates and associated confidence scores for all detected joints. These frame-wise JSON files collectively represented the temporal trajectories of individual joints across the gait cycles.

*2.3.2.2 Joint Trajectory Extraction, Joint Angle Calculation, and Data Smoothing*

The OpenPose generated JSON files were subsequently analyzed using a custom-written software pipeline designed for joint trajectory extraction and processing. This software parsed the JSON files to extract time-series joint coordinate data for all relevant anatomical landmarks.

Only knee joint angle estimation and its progression were analyzed in this study. The knee joint angle was defined as the angle between the thigh and leg segments, calculated using the hip, knee, and ankle joint coordinates.

The resulting knee angle time-series data were subsequently smoothed using a Savitzky–Golay filter to reduce high-frequency noise while preserving physiologically meaningful gait patterns.

Only smoothed joint angle data derived from complete and consistently tracked joint trajectories were included in the final analysis.

### *2.4. Statistical Analysis*

Temporal and kinematic gait variables are presented as mean and standard deviation (SD). Normality of data distribution was assessed prior to inferential analysis. Comparisons between experimental scenarios were performed using parametric tests (Student’s t-test or one-way analysis of variance [ANOVA]) for normally distributed variables, and non-parametric tests for variables that did not meet normality assumptions. All statistical analyses were conducted using GraphPad Prism (GraphPad Software, San Diego, CA, USA).

## 3. Result

Twelve participants who met all inclusion and exclusion criteria and conditions for video data analysis, generation of a completed library of Jason files, and validity for subsequent analysis are presented below. All outliers were re-visited and reassessed. However, only six participants’ joint kinematic data were fully compatible for statistical analysis. Two independent observers repeated the data extraction to determine inter-observer variation.

### 3.1. Spatiotemporal Gait Statistics

Spatiotemporal parameter measurements were solely done by manual observation of the gait distance and time for two gait cycles across the scenarios. S1(the baseline) and S2 are supposed to be less stressful compared to S3 and S4.

#### 3.1.1 Distance walked over two gait cycles, time and velocity

Mean walking distance differed significantly across scenarios ($p = 0.02$) (Fig. 3. (a, b)). Distance increased from baseline in S1 & S2 and remained elevated in S3 (239.4 ± 91.2 cm), before decreasing in S4 (204.7 ± 65 cm) (Table 1). This reflects the baseline calm walking to stress-induced acceleration of movements upon visibility of an e-scooter (S2), followed by significant slowing down (S3 & S4) under imminent collision threat.

The total time spent for two gait cycles shows the reciprocal relationship to the distance walked. Total gait cycle time differed significantly ($p = 0.04$), with a pairwise difference between S1 and S4 ($p = 0.02$) (Fig. 3. (c)). Comparison between two gait cycles indicated that stress impact occurred during the second gait cycle (Fig. 4).

The combination of increased time and reduced distance in S4 resulted in a marked reduction in walking velocity.

Step time increased slightly from 1.33 ± 0.1 s in S1 to 1.4 ± 0.3 s in S4, although these differences were not statistically significant (Fig. 3. (d)). The consistent directional increase suggests subtle cadence modulation under stress (Table 1).

Gait velocity is a collective response of distance and time. Gait velocity differed significantly across scenarios ($p = 0.04$), with a highly significant reduction between S2 and S4 ($p = 0.001$).

Velocity increased in S2 (92.6 ± 27.9 cm/s) compared with baseline (84.5 ± 18.7 cm/s), suggesting anticipatory acceleration in response to approaching the e-scooter. In contrast, velocity decreased substantially in S4 (71.6 ± 20.5 cm/s), reflecting defensive deceleration. The S2–S4 contrast represents the strongest biomechanical modulation observed in the dataset (Table 1, Fig. 3.e).

**Submitted** Table 1. Distribution of spatiotemporal parameters

| Scenario | Total Distance GC1+GC2 (cm), Mean, SD | Total Time GC1+GC2 (s), Mean, SD | Step Time (s) | Velocity (cm/s) | Gait Cycle | Time per each GC (s) Mean, SD | Stance of each GC (s) (mean, SD) | Swing of each GC (s) (Mean, SD) |
|---|---|---|---|---|---|---|---|---|
| S1 | 226.5, 53.5 | 2.67, 0.2 | 1.33, 0.1 | 84.5 | GC1 | 1.35, 0.11 | 0.92, 1.1 | 0.41, 0.03 |
| | | | | | GC2 | 1.34, 0.15 | 0.9, 0.16 | 0.37, 0.17 |
| S2 | 247, 77 | 2.67, 0.45 | 1.34, 0.22 | 92.6 | GC1 | 1.22, 0.17 | 0.81, 0.13 | 0.44, 0.16 |
| | | | | | GC2 | 1.3, 0.12 | 0.86, 0.15 | 0.46, 0.08 |
| S3 | 239.4, 91.2 | 2.8, 0.5 | 1.4, 0.22 | 85.8 | GC1 | 1.18, 0.24 | 0.81, 0.15 | 0.4, 0.11 |
| | | | | | GC2 | 1.5, 0.5 | 1.02, 0.46 | 0.39, 0.14 |
| S4 | 204.7, 65 | 2.83, 0.6 | 1.4, 0.3 | 71.6 | GC1 | 1.17, 0.30 | 0.78, 0.23 | 0.44, 0.09 |
| | | | | | GC2 | 1.47, 0.47 | 1.04, 0.46 | 0.47, 0.1 |

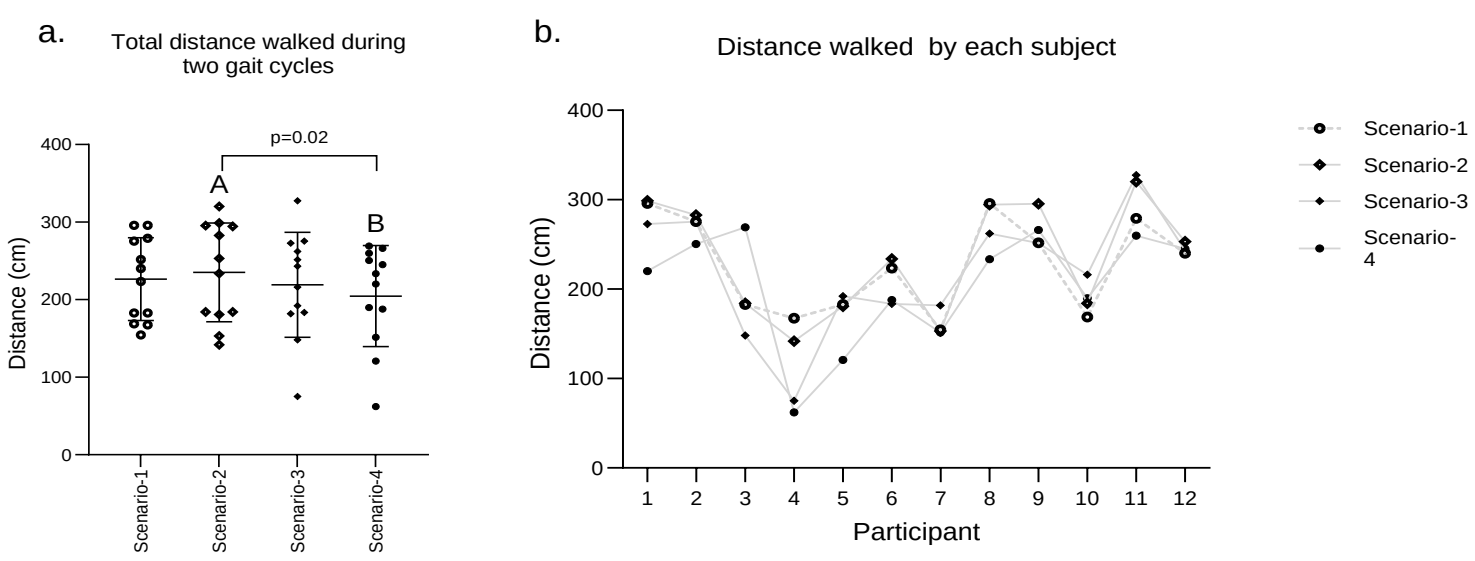

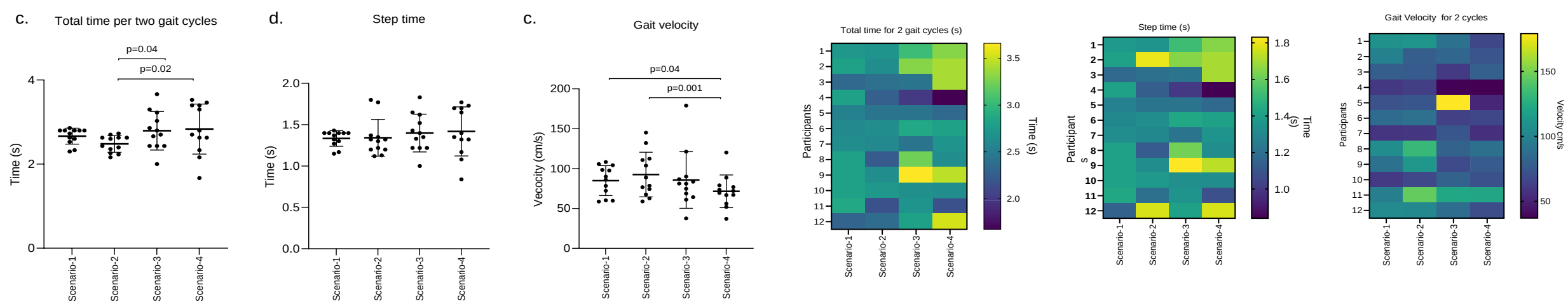

Fig. 3. Comparison of spatiotemporal parameters (a, b), Statistical comparisons of gait distance (c,d), gait time, and (e) gait velocity done using ANOVA and Student’s t-test. Heat maps reflect the qualitative data point variations; scenarios 3 and 4 indicate the widest scatter

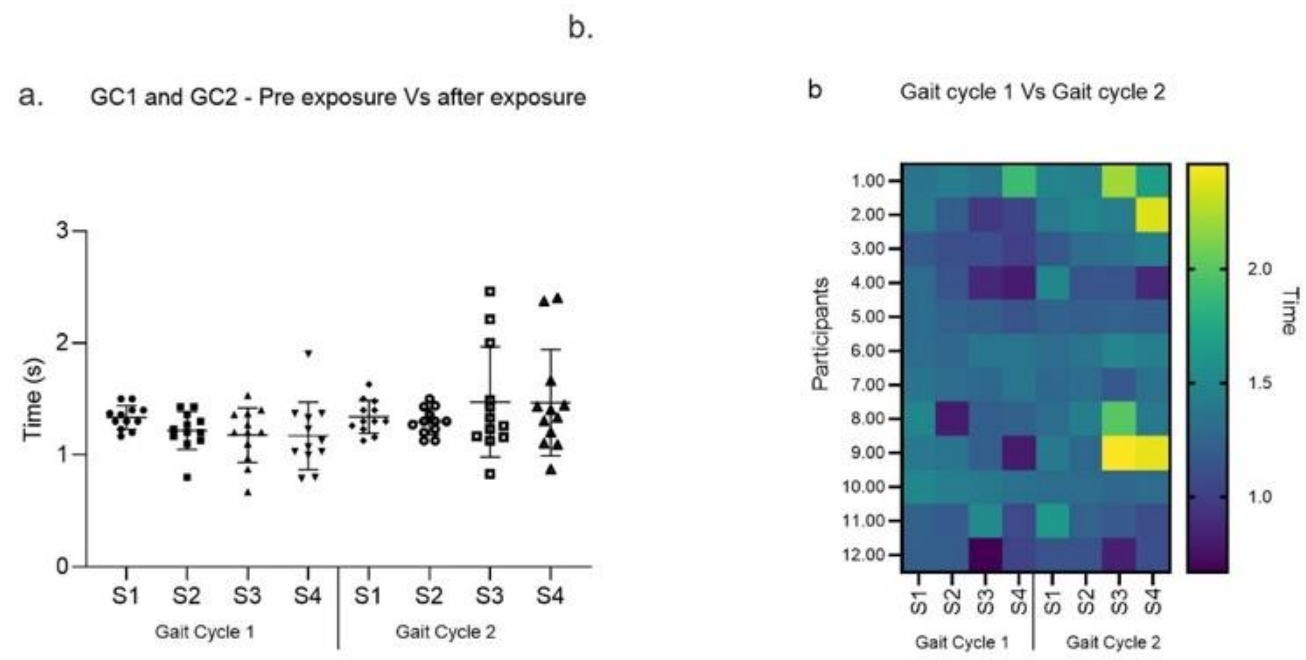

Fig. 4. Comparison of two gait cycles. The heat map indicates the widest data scatter in gate cycle 2 (GC2)

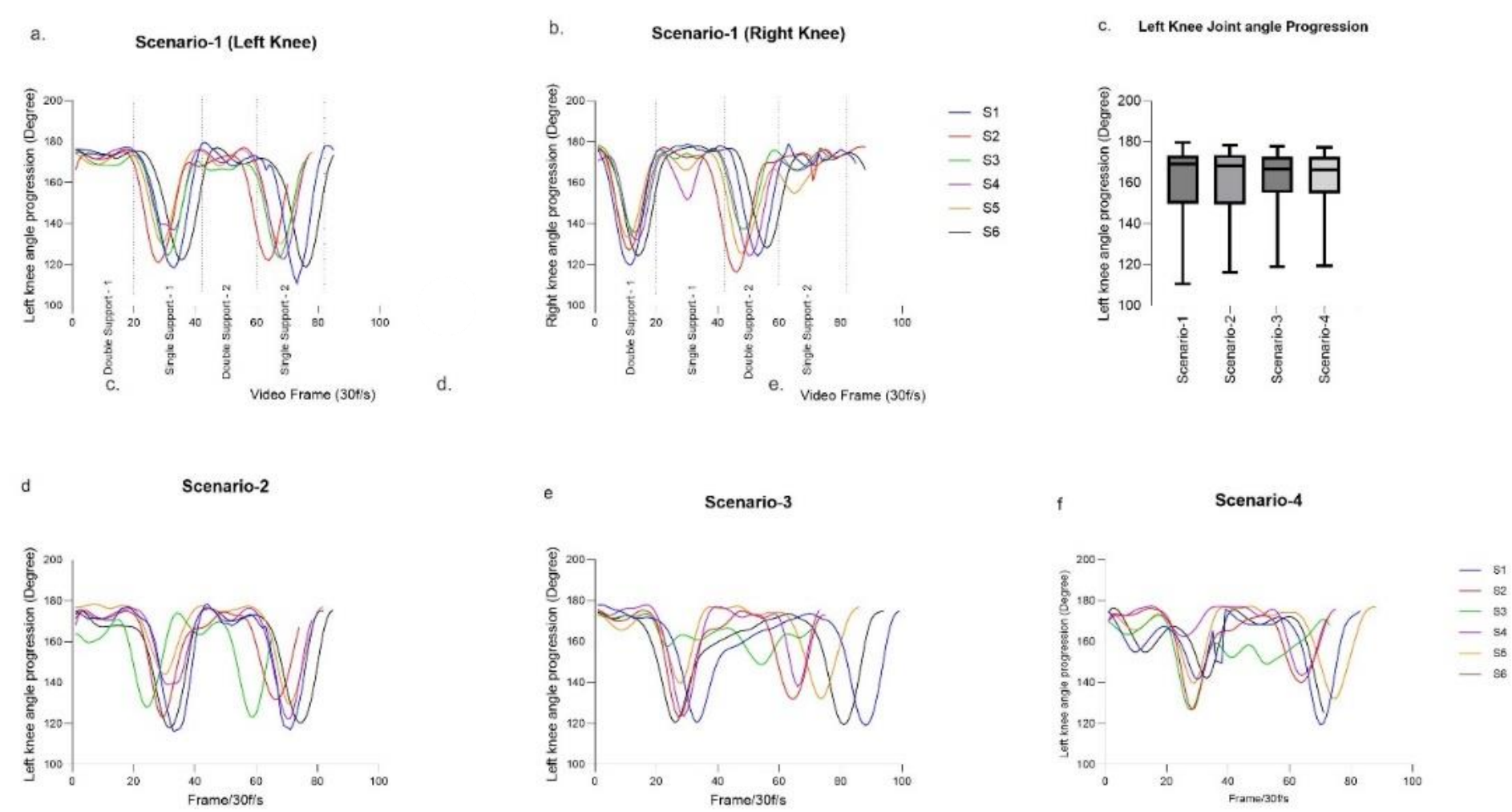

Fig. 5. The left knee joint angle progression with the comparative right knee

## *3.2. Joint Kinematics*

Fig. 5 illustrates the progression of the knee joint angle with the different phases of the gait. Fig. 5(a) and Fig. 5(b) represent the baseline, which is more consistent across the subjects. In Fig. 5(b), the right knee joint shows a slight glitter between gait phases, indicating obstruction of the limb view by the sagittal camera. The OpenPose-based reading and gait estimation for the right lower limb thus becomes less reliable.

The visual observation (i.e., qualitative analysis) also reveals the jitteriness of the angle progression in scenarios 3 and 4 (Fig. 5(d) and (e).

## 4. Discussion

This VR-simulated e-scooter encounter influences pedestrian gait and lower-limb kinematics in a controlled but realistic traffic environment. Our main findings were that e-scooter encounters were associated with measurable changes in spatiotemporal gait parameters. That is, reduced gait velocity and distance walked over two gait cycles, along with prolongation of total time and step time compared with non-stressful pedestrian-only walking. These adaptations were most pronounced in the simulated crash-in scenario, suggesting a graded and threat-dependent modulation of gait.

In addition, knee joint angle trajectories derived from 2D video and pose estimation revealed altered flexion–extension patterns across scenarios (i.e., jitteriness in knee joint angle progression), indicating that pedestrians adjust both timing and posture when sharing space with fast-approaching micro-mobility vehicles.

Our findings align with the growing literature showing that e-scooters have substantial risk to road users in urban environments. Hospital-based studies report high rates of e-scooter-related injuries, with incidence estimates exceeding those for cyclists and pedestrians in some regions. While most reports focus on riders, pedestrians are increasingly exposed to near-misses and collisions on sidewalks and shared paths, with cases of e-scooter–related pedestrian injury and craniofacial trauma documented in the emergency medicine and maxillofacial literature. (Zhao et al. 2022, Flaherty et al. 2025, Collins et al. 2026)

Our results thus extend this work by demonstrating that even in the absence of physical contact, the presence and behavior of e-scooters can induce measurable changes in pedestrian gait which is an objective marker of altered perceived safety.

The observed reductions in gait velocity and increases in step time during e-scooter encounters are consistent with cautious gait strategies described in other contexts of perceived instability or environmental threat. Experimental work on pedestrian–vehicle interactions suggests that posture, stance-phase duration, and limb-loading patterns influence both collision risk and injury severity (Li 2021).

In this regard, the prolongation of stance and step time in the crash-in scenario may reflect an attempt to stabilize the body and the readiness for an evasive action. Complementary VR studies of pedestrian–micro-mobility interactions have shown that evasive strategies and risk perception strongly depend on the relative trajectories and timing of the agents; early evasive actions by micro-vehicle users, as well as wider shared paths, can substantially reduce objective risk (Chen 2021, Li 2021, Wang 2024).

Our data support that pedestrians themselves also modulate gait to manage perceived collision risk when e-scooters approach at higher speeds or on conflicting paths.

The use of 2D sagittal-plane video combined with OpenPose-based pose estimation and subsequent smoothing provided a practical and non-invasive way to quantify knee joint kinematics in a VR laboratory setting. Previous validation studies have shown that OpenPose can estimate temporo-spatial gait parameters and sagittal-plane hip and knee angles with acceptable accuracy compared with three-dimensional marker-based motion capture systems (Ino 2024, Ge, Wu et al. 2025).

In our study setting (fixed camera position, controlled walkway, and Savitzky–Golay filtering of complete trajectories) adhered closely to recommended workflows for reliable 2D gait analysis, thereby supporting the validity of the knee-angle findings. At the same time, we noted that contralateral sagittal plane limb kinematics are less reliable for pose estimation due to the hindrance of the opposite side limb during walking. Similarly, ankle kinematics and side-by-side motions are less accurately captured with this approach, and future work may benefit from multi-camera or 3D systems when feasible.

Our VR simulation used a realistic street-and-walkway 360 ° video to mimic a real-time walking experience. Therefore, this methodological approach, while being safe, mimics a strong realism. However, the VR scenario induces non-stressful versus stressful visuals, which can cause VR-dizziness and sickness, confounding the measurement effects. In the exclusion criteria and in the post-experimental questionnaire, such experiences and any predisposing VR-related discomfort or illnesses. This has to be explicitly inquired and recorded according to QU-IRB guidelines as well.

This study has several limitations. First, the sample comprised a relatively small, homogeneous group of healthy university students, which may limit generalizability to older adults, children, or individuals with gait impairments who might respond more strongly to e-scooter encounters. Second, we analysed only two gait cycles per trial and focused primarily on sagittal-plane left knee kinematics; a more comprehensive analysis could include longer walking bouts and multi-joint coordination patterns. Third, while the VR scenarios were based on real-world recordings, actual street environments may involve additional cues (ambient noise, crowd density, complex obstacles) that further influence gait and risk perception.

Despite these limitations, our findings illustrate the potential of VR-based biomechanical assessment to capture pedestrian adaptations to emerging modes of micro-mobility. By linking specific e-scooter behaviors (speed, trajectory, conflict type) with quantifiable changes in gait, this approach can inform the design of safer shared spaces, as well as the development of e-scooter operational rules and driver assistance technologies. In addition, the workflow combining consumer-grade cameras, head-mounted displays, and open-source gait analysis tools offers a scalable template for multi-site studies and for evaluating interventions such as modified infrastructure, speed limits, or warning systems.

## 5. Conclusion

In a VR-simulated urban environment, encounters with e-scooters led to distinct, threat-dependent modifications in pedestrian gait, including reduced walking speed, longer step and cycle times, and altered knee joint kinematics compared with non-stressful pedestrian-only walking. These objective biomechanical adaptations suggest that pedestrians respond to the presence and behaviour of micro-mobility vehicles by adopting more cautious gait patterns, even in the absence of physical contact. Together with prior evidence on e-scooter injuries and pedestrian discomfort on shared facilities, our findings underscore the need for infrastructure designs and operational policies that mitigate conflict between e-scooters and pedestrians. VR-based gait analysis using pose-estimation techniques offers a promising tool for systematically testing such interventions and for advancing a more comprehensive, human-centered approach to micro-mobility safety.

## Acknowledgements

This research was supported by the Undergraduate Research Experience Program (UREP) grant from the Qatar Research, Development and Innovation Council (Grant No. QRDI - UREP31-013-3-004). The authors acknowledge the support of Qatar University, including the College of Engineering's Qatar Transportation and Traffic Safety Centre VR Laboratory, as well as the College of Medicine, QU-Health for the support of travel and conference participation. During manuscript preparation, Grammarly, ChatGPT (OpenAI), and Perplexity AI were used for language editing and formatting. All outputs were reviewed, and the authors take full responsibility for the content.

## References

Cao, Z., G. Hidalgo, T. Simon, S. E. Wei and Y. Sheikh (2021). "OpenPose: Realtime Multi-Person 2D Pose Estimation using Part Affinity Fields." IEEE Transactions on Pattern Analysis and Machine Intelligence 43(1): 172–186.

Chen, J. (2021). "Influence of gait stance and vehicle type on pedestrian injury biomechanics." Journal of Biomechanical Engineering.

Collins, A. S., N. A. Mirsky, R. Mangal, S. Prasad and S. R. Thaller (2026). "Craniofacial Injuries Associated With Micromobility Vehicles: A Retrospective Analysis Using the NEISS Database (2014-2023)." J Craniofac Surg 37(3-4): e136–e139.

Flaherty, M. R., A. Rosenfield and D. Rosenfield (2025). "Pediatric Injuries Among Those Using Active Transportation & Micromobility Devices." Pediatr Clin North Am 72(6): 1097–1109.

Fu, T., W. C. Hu, L. Miranda-Moreno and N. Saunier (2019). "Investigating secondary pedestrian-vehicle interactions at non-signalized intersections using vision-based trajectory data." Transportation Research Part C-Emerging Technologies 105: 222–240.

Ge, F., C. Wu and S. Xu (2025). "Reliability and Validity of OpenPose for Measuring Hip-Knee-Ankle Angle." Scientific Reports.

Ino, T. (2024). "Validity and Reliability of OpenPose-Based Motion Analysis." Sensors.

Li, X. (2021). "Kinematic analysis of pedestrian behavior in virtual reality environments." Frontiers in Bioengineering and Biotechnology.

Merchant, H. and A. P. Georgopoulos (2006). "Neurophysiology of perceptual and motor aspects of interception." J Neurophysiol 95(1): 1–13.

Ni, Y., M. L. Wang, J. Sun and K. P. Li (2016). "Evaluation of pedestrian safety at intersections: A theoretical framework based on pedestrian-vehicle interaction patterns." Accident Analysis and Prevention 96: 118–129.

Pulido-Valdeolivas, I., D. Gomez-Andres, J. A. Martin-Gonzalo, J. Lopez-Lopez, E. Gomez-Barrena, J. J. Sanchez Hernandez and E. Rausell (2013). "Gait parameters in a reference sample of healthy Spanish schoolchildren: multivariate descriptive statistics and asymmetries observed in left and right cycles." Neurologia 28(3): 145–152.

Sethi, D., S. Bharti and C. Prakash (2022). "A comprehensive survey on gait analysis: History, parameters, approaches, pose estimation, and future work." Artif Intell Med 129: 102314.

Wang, Y. (2024). "Biomechanical factors influencing pedestrian injury severity in vehicle collisions." Accident Analysis and Prevention.

Zhao, Y., J. Cao, Y. Ma, S. Mubarik, J. Bai, D. Yang, K. Wang and C. Yu (2022). "Demographics of road injuries and micromobility injuries among China, India, Japan, and the United States population: evidence from an age-period-cohort analysis." BMC Public Health 22(1): 760.